\documentclass[12pt]{article}
\usepackage{epsfig}
\usepackage{graphicx}
\usepackage{amsmath}
\usepackage{amssymb}
\usepackage[table]{xcolor}
\usepackage{tikz,xcolor,hyperref}
\usepackage{lipsum}
\usepackage[font=scriptsize,labelfont=bf]{caption}
\usepackage{datetime}
\usepackage[font=scriptsize,labelfont=bf]{caption}
\usepackage{datetime}
\usepackage{tikz}
\usepackage{subfig}
\usepackage{makecell}
\usepackage{multirow}

\definecolor{lime}{HTML}{A6CE39}
\DeclareRobustCommand{\orcidicon}{%
	\begin{tikzpicture}
		\draw[lime, fill=lime] (0,0)
		circle [radius=0.16]
		node[white] {{\fontfamily{qag}\selectfont \tiny ID}};
		\draw[white, fill=white] (-0.0625,0.095)
		circle [radius=0.007];
	\end{tikzpicture}
	\hspace{-2mm}
}

\foreach \x in {A, ..., Z}{%
	\expandafter\xdef\csname orcid\x\endcsname{\noexpand\href{https://orcid.org/\csname orcidauthor\x\endcsname}{\noexpand\orcidicon}}
}

\begin{document}
\begin{center}
\Large{\bf Black Hole Solutions with Constant Ricci Scalar in a Model of Finsler Gravity}\\
\small \vspace{1cm}{\bf Z. Nekouee \orcidB{} $^{\ddagger\dagger}$
\footnote{zohrehnekouee@gmail.com}}, {\bf S. K. Narasimhamurthy \orcidA{} $^{\dagger}$
\footnote{nmurthysk@gmail.com ~~~(Corresponding author)}} and {\bf S. K. J. Pacif \orcidC{} $^{*}$ \footnote{shibesh.math@gmail.com}}\\
\vspace{0.5cm}$^{\ddagger}${\it School of Physics, Damghan University, Damghan, 3671641167, Iran.}\\
\vspace{0.5cm}$^{\dagger}${\it Department of PG Studies and Research in Mathematics, Kuvempu University,\\ Jnana Sahyadri, Shankaraghatta-577 451, Shivamogga, Karnataka, India.}\\
\vspace{0.5cm}$^{*}${\it Pacif Institute of Cosmology and Selfology (PICS),\\ Sagara, Sambalpur-768 224, Odisha, India.}\\
\end{center}\vspace{1.2cm}
\date{}
\begin{abstract}
 Ricci scalar being zero is equivalent to the vacuum field equation in Finsler space-time. The Schwarzschild metric can be concluded from the field equation's solution if the space-time conserves spherical symmetry. This research aims to investigate Finslerian Schwarzschild-de Sitter space-time. Recent studies based on Finslerian space-time geometric models are becoming more prevalent because the local anisotropic structure of space-time influences the gravitational field and gives rise to modified cosmological relations. We suggest a gravitational field equation with a non-zero cosmological constant in Finslerian geometry and apprehend that the presented Finslerian gravitational field equation corresponds to the non-zero Ricci scalar. In Finsler geometry, the peer of spherical symmetry is the Finslerian sphere. Assuming space-time to conserve the "Finslerian sphere" symmetry, the counterpart of the Riemannian sphere (Finslerian sphere) must have a constant flag curvature ($\lambda$). It is demonstrated that the Finslerian covariant derivative of the geometric part of the gravitational field equation is preserved under a condition using the Chern connection.
According to the string theory, string clouds can be defined as a pool of strings made due to symmetry breaking in the universe's early stages. We find that for $\lambda\neq1$, this solution resembles a black hole surrounded by a cloud of strings.
Furthermore, we investigate null and time-like geodesics for $\lambda=1$. In this regard, the photon geodesics are obtained that are the closest paths to the photon sphere of the first photons visible at the black hole shadow limit. Also, circular orbit conditions are obtained for the effective potential.\\

\end{abstract}
\newpage
\section{Introduction}\label{sec:intro}
In 1915, Albert Einstein proposed general relativity (GR), which describes gravity as a sign of the curvature of space-time \cite{Faber}. Space-time geometry affects matter, and matter also determines geometry. Einstein's field equations relate the metric coefficients to the matter distribution. A detailed discussion of the exact solutions of Einstein's field equations can be found in \cite{Negi,HMManjunath}.
Up to the present, one of the most prosperous gravity theories is Einstein's general relativity because it coincides with experiments of the highest achievable accuracy, like Mercury perihelion precession, light bending, radar echo delay, gravitational redshift, etc.
Nevertheless, in Einstein's theory of gravity, there may be problems in correctly apprehending gravity at all scales, including it cannot accurately express the universe's accelerated expansion on large scales, and it is very tough to develop a complete theory of quantum gravity on micro scales.
This led to assorted modifications of gravity theory, such as superstring theory \cite{Green}, scalar-tensor theories \cite{Brans}, Lovelock gravity \cite{Lovelock}, H$\check{o}$rava-Lifshitz gravity \cite{Horava}, f(R) gravity \cite{Souza}, and so on.\\

As we know, Riemannian geometry describes the space-time of Einstein's general relativity. Since GR has problems, scientists have suggested modifying the theory of gravity by improving the differential geometry.
As a generalization of Riemannian geometry, Finsler geometry \cite{Bao2,Roxburgh} is the most general geometry where the line element depends on space-time coordinates and tangent vectors. $d\mathfrak{s}^2=\mathfrak{f}(x^i, dx^j)$;
$d\mathfrak{s}^2=\mathfrak{f}(x^i,\lambda dx^j)=\lambda^2 \mathfrak{f}(x^i, dx^j)$, and the metric can also be given by $d\mathfrak{s}^2= \mathfrak{G}_{ij}dx^idx^j$, where noting that $\mathfrak{G}_{ij}$ is related to $dx^j$. Chern \cite{SChern} stated that under quadratic restriction, i.e., when $\mathfrak{f}(x^i, dx^j)$ transforms into $dx^j$ in quadratic form, Finslerian geometry coincides with Riemannian geometry. At every point in the Riemannian manifold, the tangent manifold is isometric to the Minkowski space-time. While at distinguished points in the Finsler manifold, the tangent manifolds are not isometric to each other \cite{Shen4}.\\
Finsler geometry describes different symmetries of space-time. Kostelecký \cite{Kostelecky} described that Lorentz invariance violation (LIV) is closely related to the Finsler geometry. A generalization of the modified Newtonian dynamical theory (MOND) in Finsler space \cite{ZChang} is used to investigate the experimental observations of Bullet Clusters. The observational data \cite{Markevitch} is compatible with the MOND theory in the Finslerian background with the quadrupole effects. Bullet Cluster convergence $k$-map shows that the gravitational force has an anisotropic distribution. So, the quadrupole contribution is necessary together with the monopole contribution to explain the experimental observations. It is crucial to note that the metric tensor of the Finsler space is a function of the position and a function of the directional argument that induces anisotropy. The quadrupole and monopole contributions explain the anisotropic and are regarded in the Finslerian gravity. Unlike Riemannian geometry, Finslerian geometry is direction-dependent, so It is a more suitable tool to settle problems related to the nature of gravity.\\
Pseudo-Finsler geometry acts as the physical geometry of space-time \cite{CPfeifer}. The potential applications of the Finslerian geometry justify a wide range of astrophysical phenomena. Nowadays, the wormhole models are also analyzed in the framework of Finsler geometric  \cite{Rahaman,Manju}.\\

Using Finslerian geometry, Bekenstein debated the relationship between gravitational and physical geometries for gravitational theory in the two-geometry paradigm \cite{Bekenstein}. S.I. Vacaru presented an introduction to Finsler-Lagrange geometry in Einstein and string gravity \cite{Vacaru1}. In Ref. \cite{Mavromatos}, Mavromatos expressed the relation between Finsler geometry and some techniques of the quantum gravity problem. Recently, considering the Finsler geometry can explain the Planckian structure of the configuration space of relativistic particles. The generalization of the obtained results is seen in Ref. \cite{Girelli}. Lobo, Loret, and Nette investigated Planck-scale-deformed relativity in the Finsler geometry framework \cite{Lobo}. Furthermore, Researchers have developed several directions for the construction of different classes of Finsler black hole solutions \cite{Rutz,Vacaru4,Vacaru5}.\\
In the last two decades, with the discovery and deep investigation of cosmic abnormalities, researchers have uncovered that the universe may be anisotropic at specified scales. Several physicists attempted to present Finslerian geometry to create a background model of the universe \cite{Bartelmann,Caponio}.
Finslerian geometry, as a natural generalization of Riemannian geometry, can provide new insight into modern physics. Inspired by these endeavors and insights, physicists have successfully described unusual phenomena.
For example, the flat rotation curves of spiral galaxies can be concluded naturally without citing dark matter \cite{SSChern}, the secular trend in the astronomical unit and abnormal variation of the secular eccentricity of the Moon's orbit could be explained by the effect of the length change of unit circle in Finsler geometry \cite{XLi}.\\
Though the theory of gravity in the framework of Finslerian geometry still faces many problems, the developments mentioned above have shown that the Finslerian theory of gravity has some potential to solve problems in GR. This line of thinking is continually evolving.\\

Compressing matter into a small space creates strong gravity, thereby eliminating the possibility of particles or radiation escaping. It is what we now call a black hole (BH), and it is classified into three categories:
\begin{itemize}
\item [1.] BHs of stellar masses that formed after the death of massive stars.
\item [2.]	Super-massive BHs with masses up to $109M\odot$, where $M\odot=2\times1033$mg is the mass of the Solar. They are located in the centers of galaxies.
\item [3.]	Primordial BHs that their presence is imputed to large-scale inhomogeneities in the nascent expansion of the universe \cite{Novikov}.
\end{itemize}
The theory that best explains gravitational phenomena at great distances is unquestionably the theory of general relativity (TGR). The concept of BHs grew out of the TGR presented in 1915. In 1916, German astronomer Karl Schwarzschild found a solution to TGR, which indicated a spherical BH, with the assumption that angular momentum, the mass's electric charge, and the universal cosmological constant are all equal to zero. It has singularities at $r=0$ and $r=2M$. But $r=2M$ is a coordinate singularity that could be removed by transforming the metric to retarded or advanced time coordinate.\\
Schwarzschild-de Sitter space-time \cite{Rindler} is an Einstein's field equations solution with a non-zero cosmological constant. Because of a symmetric spherical gravitational source, it is also equivalent to the gravitational field. Nowadays, many geometers and physicists perused the famous Schwarzschild–de Sitter solution and debated black holes (BHs) in the (Anti-)de Sitter space-time and have investigated the Finsler-Randers cosmological models \cite{Raushan,Papagiannopoulos,Basilakos,Chaubey}.\\
A BH is a structure that plays a significant role in the theory of gravity and helps us understand some of the thermodynamic aspects and potential formulations of quantum gravity.
Therefore, others discovered the BH solutions in other theories of gravity, among which exact solutions in the context of Lovelock gravity were discovered \cite{Cai,Hennigar,Myers}. Letelier first conducted a theoretical analysis and got the Einstein equations solutions corresponding to a BH surrounded by a cloud of strings \cite{Letelier}.
The original concept of string theory was extended to include considering a cloud of strings and investigating any potentially measurable effects on the long-range gravitational fields of a BH.\\
The successful verification of TGR was mainly thanks to the geodesics of Schwarzschild. There has been a growing interest in exploring the geodesic behavior around BHs in the last few years \cite{Al Zahrani,Frolov,Abdujabbarov,Ali}.
 Because measuring gravitational waves is so challenging, geodetic studies are the only practical method for physicists. Schwarzschild geodesic describes particle motion test in Schwarzschild metric or the motion in the gravitational field of a fixed central mass M. On the other hand, to understand the geometric structure of space-time near the BH, one can use the motion of a particle in its vicinity. In a pseudo-Riemannian manifold, a geodesic is a generalization of the concept of a straight line in Euclidean space. Its tangent vector is always in the same direction and is the shortest path between two points.
Time-like geodesics are curves with extreme lengths between two points and represent the path of massive particles, whereas null geodesics indicate the photon's path.

The current work expands the classical Schwarzschild–de Sitter solution to the Finsler framework. Therefore, it demonstrates a deviation in geometric structure, consequently in gravitation theory from the Riemannian framework. The Finslerian sphere plays a crucial role in the classical Schwarzschild-de Sitter metric perturbation.
 Finslerian sphere is the counterpart of spherical symmetry in Finsler geometry.
To describe the properties of most celestial bodies, the Finsler sphere should be like a sphere.
In mathematics, it should be topologically similar to a sphere. In addition, the Finsler sphere should maintain symmetry as much as possible. The proved theorem in Ref. \cite{Wang} shows that the two-dimensional Finsler space has just one independent Killing vector. So, the Finslerian sphere has also the same property.\\

In Finsler manifolds, the Ricci curvature is crucial \cite{BaoR2, Robles}. It is described as the Riemann curvature's trace on each tangent space.
If you think of a Riemannian manifold as a white egg and a Finsler manifold as an Easter egg, Finslerian manifolds are far more colorful. The color and its rate of change over the manifold are described by several non-Riemannain quantities, including the mean Cartan torsion $C$, the mean Landsberg curvature $L$, and the S-curvature $S$. (see \cite{Shen4})
Studying and describing Finsler metrics with constant flag curvature or constant Ricci curvature is one of the crucial issues in Finsler geometry. Einstein metrics are the answer to the GR's Einstein's Field equations in Finsler geometry. It is critical to calculate Riemann curvature and Ricci curvature for Einstein-Finsler $(\alpha, \beta)$ -metrics to describe them \cite{Narasim}.

Finsler and Finsler-like geometries, which are metric generalizations of Riemannian geometry and depend on location, velocity/momentum/scalar coordinates, are potential metric geometries that can intrinsically explain the motion.
These dynamic geometries can explain locally anisotropic phenomena, Lorentz violations, field equations, FRW and Raychaudhuri equations, geodesics, and effects of dark matter and dark energy \cite{Kostelecky,4,9,15,16,17,18,20}, among other things.
Taking this approach, one can interpret the gravitational field as a metric of a generalized space-time and see how the field creates a force that constrains motion.
This scenario demonstrates space-time's Finslerian geometrical nature. Numerous studies in various geometrical fields and physical structures have advanced research for theoretical and observational approaches during the past few years in the context of Finsler geometry applications. We cite a few publications from the literature on Finsler geometry applications \cite{21,24}.

The Finslerian Schwarzschild BH is a warped product of space-time with constant Finslerian curvature in its two-dimensional subspace. Four constants of motion have been deduced from the geodesic equations of Finslerian Schwarzschild space-time from its unique geometrical structure.
Both orbital precession and orbital plane precession are impacted by the Finslerian parameter $\varepsilon$, which characterizes the difference between Finslerian Schwarzschild BH and Schwarzschild BH \cite{Xin}.\\
Based on the above study, in this paper, we have investigated the geodesics in Finslerian BH, and the paper is organized as follows:
In section~\ref{sec2}, we quickly review the concepts of Finslerian geometry. In section~\ref{sec:2}, we have obtained a solution for the field equations in the Finslerian space-time. Section~\ref{sec:3} is divided into two subsections. In subsection~\ref{subsec:31}, we have discussed the null geodesics of the Finslerian space-time, and in subsection ~\ref{subsec:32}, we have considered time-like geodesics and analyzed the effective potential. Finally, we conclude our article in section~\ref{sec:4}.
\section{Preliminaries and Notations of Finsler Geometry}\label{sec2}
Finsler structure on manifold $M$ is defined as function $\mathcal{F}: TM\rightarrow [0, \infty)$  which satisfies the below properties:\\
1. Regularity: $\mathcal{F}$ is smooth function on the $TM\backslash\{0\}$.\\
2. Positive Homogeneity: $\mathcal{F}(x, cy)= c\mathcal{F}(x, y)$ for all $c>0$.\\
3. Strong Convexity: The $n\times n$ Hessian matrix
\begin{equation}\label{Eq3}
g_{\upsilon\omega}=\frac{\partial^{2}\bigg(\frac{\mathcal{F}^{2}}{2}\bigg)}{\partial y^{\upsilon}\partial y^{\omega}}.
\end{equation}
In physics, the Finsler structure $\mathcal{F}$ is not positive-definite at every point of the Finsler manifold, and it is the function of ($x^{i}, y^{i}$) that $y^{i}\equiv\frac{dx^{i}}{d\tau}$ represent velocity \cite{Li1}. The pair ($M, \mathcal{F}$) is called Finsler space. We concentrate on studying Finsler space-time with Lorentz's signature.\\
A Finslerian metric is referred to as Riemannian if $\mathcal{F}^{2}$ is quadratic in $y$. For the Finsler manifold, the geodesic equation is as follows:
\begin{equation}\label{Eq1}
\frac{d^{2}x^{\upsilon}}{d\tau^{2}}+2\mathcal{G}^{\upsilon}(x,y)=0,
\end{equation}
where the geodesic spray is
\begin{equation}\label{Eq2}
\mathcal{G}^{\upsilon}=\frac{1}{4}g^{\upsilon\omega}\bigg(\frac{\partial^{2}\mathcal{F}^{2}}{\partial x^{k}\partial y^{\omega}}y^{k}-\frac{\partial \mathcal{F}^{2}}{\partial x^{\omega}}\bigg).
\end{equation}
From the geodesic \eqref{Eq1}, it can be proven that the Finslerian structure $\mathcal{F}(x,\frac{dx}{d\tau})$ along the geodesic is constant \cite{Li1,SK}.\\
 There is a geometric invariant in Finsler geometry called the Ricci scalar, which is represented by "$Ric$" and has the following form \cite{Bao}
\begin{equation}\label{Eq4}
Ric=R^{\mu}_{\mu}=\frac{1}{\mathcal{F}^{2}}\bigg(2\frac{\partial \mathcal{G}^{\mu}}{\partial x^{\mu}}-y^{\nu}\frac{\partial^{2} \mathcal{G}^{\mu}}{\partial x^{\nu}\partial y^{\mu}}+2\mathcal{G}^{\nu}\frac{\partial^{2} \mathcal{G}^{\mu}}{\partial y^{\nu}\partial y^{\mu}}-\frac{\partial \mathcal{G}^{\mu}}{\partial y^{\nu}}\frac{\partial \mathcal{G}^{\nu}}{\partial y^{\mu}}\bigg),
\end{equation}
where $Ric$ exclusively relies on the Finsler structure and is insensitive to connections.\\
For a tangent plane $\Xi \subset T_{x}M$ and a vector $0\neq y \in T_{x}M$, the flag curvature is defined as
\begin{equation}\label{Eq4.1}
K(\Xi,y)\equiv \frac{g_{\lambda\mu}R^{\mu}_{\nu}u^{\nu}u^{\lambda}}{\mathcal{F}^{2}g_{\rho\vartheta}u^{\rho}u^{\vartheta}-
(g_{\sigma\kappa}y^{\sigma}u^{\kappa})^{2}},
\end{equation}
where $u\in \Xi$.  The generalization of sectional curvature in Riemannian geometry is flag curvature, which is a geometrical invariant in Finsler geometry.  The Ricci scalar is the trace of $R^{\mu}_{\nu}$. It is the predecessor of flag curvature. Thus, under the coordinate transformation, the value of the Ricci scalar is invariant.\\
In Finslerian geometry, the Ricci tensor is given by
\begin{equation}\label{Eq5}
Ric_{\upsilon\omega}=\frac{\partial^{2}}{\partial y^{\upsilon}y^{\omega}}\bigg(\frac{1}{2}\mathcal{F}^{2}Ric\bigg),
\end{equation}
which was first proposed by Akbar-Zadeh \cite{Akbar-Zadeh}. It plays an essential role in investigating the Finsler-Einstein structures.\\
Finslerian modified formula for scalar curvature is
\begin{equation}\label{Eq5.1}
	\mathcal{S}=g^{\mu\nu}Ric_{\mu\nu},
\end{equation}
and Einstein tensor formula is
\begin{equation}\label{Eq5.2}
	\mathcal{G}_{\mu\nu}=Ric_{\mu\nu}-\frac{1}{2}g_{\mu\nu}\mathcal{S}.
\end{equation}
Since it is derived from $Ric$, connections do not affect the Finslerian-modified Einstein tensor. It is solely reliant on the Finslerian structure. As a result, the Finslerian gravitational field equations are also insensitive to the connections.\\
Li and Chang \cite{Li1} presented the following Einstein field equation in Finsler space-time,
\begin{equation}\label{Eq5.3}
\mathcal{G}^{i}_{j}=8\pi_{\mathcal{F}}G\mathcal{T}^{i}_{j},
\end{equation}
where $\mathcal{T}^{i}_{j}$ and $\mathcal{G}^{i}_{j}$ are the energy-momentum tensor and modified Einstein tensor, respectively.
Stavrinos \cite{PStavrinos} presented the following Einstein-type equation in Finsler space-time with considering $\Lambda \neq 0$ as follows form,
\begin{equation}\label{Eq5.4}
Ric^{ij}-\frac{1}{2}a^{ij}\mathcal{S}-\Lambda a^{ij} =-8\pi G\mathcal{T}^{ij},
\end{equation}
where $Ric_{ij}$, $a_{ij}$, $\mathcal{T}_{ij}$, $\mathcal{S}$, and $\Lambda$ denote
the Ricci tensor, osculating Riemannian metric, the energy-momentum tensor related with $a_{ij}$, the scalar curvature, and the cosmological constant, respectively.\\
The metric of the Riemannian sphere is given by
\begin{equation}\label{Eq5.5}
F_{RS}=\sqrt{y^{\theta}y^{\theta}+\sin^{2}(\theta)y^{\phi}y^{\phi}},
\end{equation}
The Riemannian sphere has spherical symmetry. Its counterpart in Finslerian geometry is called the Finslerian sphere. It is topologically analogous to a two-dimensional sphere, which has a constant Ricci scalar and
a constant flag curvature. It is a two-dimensional Randers–Finsler space. The metric for the Finslerian sphere is written as follows \cite{Li1,Bao1},
\begin{equation}\label{Eq5.6}
\mathcal{F}_{FS}=\frac{\sqrt{\bigg(1-\varepsilon^{2}\sin^{2}(\theta)\bigg)y^{\theta}y^{\theta}+\sin^{2}(\theta)y^{\phi}y^{\phi}}}{1-\varepsilon^{2}\sin^{2}(\theta)}-
\frac{\varepsilon\sin^{2}(\theta)y^{\phi}}{1-\varepsilon^{2}\sin^{2}(\theta)},
\end{equation}
where $0\leq\varepsilon<1$ is named the Finslerian parameter.  If $\varepsilon=0$, the Finslerian sphere declines to the Riemannian sphere. Also, the metric \eqref{Eq5.6} is non-reversible for $\varphi\rightarrow-\varphi$. The volume of the Finslerian sphere is $4\pi$ corresponding to the Busemann–Hausdorff volume form \cite{Shen4}.
\section{Finslerian black hole solution}\label{sec:2}
Now, we consider the Finsler metric $\mathcal{F}=\mathcal{F}(x, y)$ as a function of $(x^{i}, y^{i})$ in a standard coordinate system that the angular coordinate is given in the following ansatz as $\bar{\mathcal{F}}^{2}(\theta, \phi, y^{\theta}, y^{\phi})$. $\bar{\mathcal{F}}$ does not depend on $y^{t}$ and $y^{r}$.
\begin{equation}\label{Eq6}
\mathcal{F}^{2}=\mathcal{B}(t,r)y^{t}y^{t}-\mathcal{A}(t,r)y^{r}y^{r}-r^{2}\bar{\mathcal{F}}^{2}(\theta,\phi,y^{\theta},y^{\phi}),
\end{equation}
therefore, Finsler metric coefficients can be obtained as follows,
\begin{align}\label{Eq7}
g_{\mu\nu}&=diag(\mathcal{B},-\mathcal{A},-r^{2}\bar{g}_{ij}),\\
g^{\mu\nu}&=diag(\mathcal{B}^{-1},-\mathcal{A}^{-1},-r^{-2}\bar{g}^{ij}),\label{Eq8}
\end{align}
where $\bar{g}_{ij}$ and its reverse are the metric that derived from $\bar{\mathcal{F}}$ and the index $i, j$ run over
angular coordinate $\theta, \phi$. By inserting the Finsler structure \eqref{Eq6} into the formula \eqref{Eq2}, we have
\begin{align}\label{Eq9}
\mathcal{G}^{t}&=\frac{\mathcal{B}^{\prime}}{2\mathcal{B}}y^{t}y^{r}+\frac{1}{4}\frac{\dot{\mathcal{B}}}{\mathcal{B}}y^{t}y^{t}+\frac{1}{4}\frac{\dot{\mathcal{A}}}{\mathcal{B}}y^{r}y^{r},\nonumber\\
\mathcal{G}^{r}&=\frac{1}{2}\frac{\dot{\mathcal{A}}}{\mathcal{A}}y^{t}y^{r}+\frac{\mathcal{A}^{\prime}}{4\mathcal{A}}y^{r}y^{r}+\frac{\mathcal{B}^{\prime}}{4\mathcal{A}}y^{t}y^{t}-\frac{r}{2\mathcal{A}}\bar{\mathcal{F}}^{2},\nonumber\\
\mathcal{G}^{\mu}&=\frac{1}{r}y^{\mu}y^{r}+\bar{\mathcal{G}}^{\mu}, ~~~ (\mu=\theta, \phi),
\end{align}
where $\bar{\mathcal{G}}^{\mu}$ is the geodesic spray coefficient derived by $\bar{\mathcal{F}}$, and the prime and dot represent the derivative with regard to $r$ and $t$ respectively. Substituting the geodesic coefficients \eqref{Eq9} in \eqref{Eq4}, we have
\begin{align}\label{Eq10}
\mathcal{F}^{2}Ric &=\bigg[\frac{\mathcal{B}^{\prime\prime}}{2\mathcal{A}}-\frac{\mathcal{B}^{\prime}}{4\mathcal{A}}\bigg(\frac{\mathcal{A}^{\prime}}{\mathcal{A}}+\frac{\mathcal{B}^{\prime}}{\mathcal{B}}\bigg)
+\frac{\mathcal{B}^{\prime}}{r\mathcal{A}}-\frac{\ddot{\mathcal{A}}}{2\mathcal{A}}+\frac{\dot{\mathcal{A}}}{4\mathcal{A}}\bigg(\frac{\dot{\mathcal{A}}}{\mathcal{A}}+\frac{\dot{\mathcal{B}}}{B}\bigg)\bigg]y^{t}y^{t}
\nonumber\\&+\bigg[\frac{-\mathcal{B}^{\prime\prime}}{2\mathcal{B}}+\frac{\mathcal{B}^{\prime}}{4\mathcal{B}}\bigg(\frac{\mathcal{A}^{\prime}}{\mathcal{A}}+\frac{\mathcal{B}^{\prime}}{\mathcal{B}}\bigg)
+\frac{\mathcal{A}^{\prime}}{r\mathcal{A}}+\frac{\ddot{\mathcal{A}}}{2\mathcal{B}}-\frac{\dot{\mathcal{A}}}{4\mathcal{B}}\bigg(\frac{\dot{\mathcal{A}}}{\mathcal{A}}+\frac{\dot{\mathcal{B}}}{B}\bigg)\bigg]y^{r}y^{r}
+\frac{2\dot{\mathcal{A}}}{r\mathcal{A}}y^{t}y^{r}\nonumber\\&+\bigg[\bar{R}ic-\frac{1}{\mathcal{A}}+\frac{r}{2\mathcal{A}}\bigg(\frac{\mathcal{A}^{\prime}}{\mathcal{A}}-\frac{\mathcal{B}^{\prime}}{\mathcal{B}}\bigg)\bigg]\bar{\mathcal{F}}^{2}.
\end{align}
Schwarzschild-de Sitter space-time is the solution of Einstein's field equations with a non-zero cosmological constant. It is equivalent to the gravitational field due to a spherically symmetric gravitational source. Nowadays, the study of the famous Schwarzschild-de Sitter solution and the debate about BHs in (anti-)de Sitter space-time is of interest to many physicists and geometers \cite{Rindler}.
We assume that $Ric$ is equal to the constant. Here, we consider $Ric=-\Lambda$, where $\Lambda$ is the cosmological constant. According to this assumption, we have equations as follows:
\begin{equation}
\bigg[\frac{\mathcal{B}^{\prime\prime}}{2\mathcal{A}}-\frac{\mathcal{B}^{\prime}}{4\mathcal{A}}\bigg(\frac{\mathcal{A}^{\prime}}{\mathcal{A}}+\frac{\mathcal{B}^{\prime}}{\mathcal{B}}\bigg)
+\frac{\mathcal{B}^{\prime}}{r\mathcal{A}}-\frac{\ddot{\mathcal{A}}}{2\mathcal{A}}+\frac{\dot{\mathcal{A}}}{4\mathcal{A}}\bigg(\frac{\dot{\mathcal{A}}}{\mathcal{A}}+\frac{\dot{\mathcal{B}}}{B}\bigg)\bigg]=-\Lambda\mathcal{B},
\end{equation}
\begin{equation}
\bigg[\frac{-\mathcal{B}^{\prime\prime}}{2\mathcal{B}}+\frac{\mathcal{B}^{\prime}}{4\mathcal{B}}\bigg(\frac{\mathcal{A}^{\prime}}{\mathcal{A}}+\frac{\mathcal{B}^{\prime}}{\mathcal{B}}\bigg)
+\frac{\mathcal{A}^{\prime}}{r\mathcal{A}}+\frac{\ddot{\mathcal{A}}}{2\mathcal{B}}-\frac{\dot{\mathcal{A}}}{4\mathcal{B}}\bigg(\frac{\dot{\mathcal{A}}}{\mathcal{A}}+\frac{\dot{\mathcal{B}}}{B}\bigg)\bigg]=\Lambda\mathcal{A},
\end{equation}
\begin{equation}\label{eq:2.13}
\bigg[\bar{R}ic-\frac{1}{\mathcal{A}}+\frac{r}{2\mathcal{A}}\bigg(\frac{\mathcal{A}^{\prime}}{\mathcal{A}}-\frac{\mathcal{B}^{\prime}}{\mathcal{B}}\bigg)\bigg]=\Lambda r^{2},
\end{equation}
\begin{equation}
\frac{2\dot{\mathcal{A}}}{r\mathcal{A}}=0.
\end{equation}

As we know that $\bar{R}ic$ is independent of $t$ and $r$, to obtain the value of $\mathcal{A}(t,r)$ and $\mathcal{B}(t,r)$, $\bar{R}ic$ must be constant (See \eqref{eq:2.13}). It requires that the two-dimensional Finsler space $\bar{\mathcal{F}}$ has constant flag curvature. The generalization of sectional curvature in Riemannian geometry is the flag curvature in Finslerian geometry. Here, we label the constant flag curvature to be $\lambda$. Hence, $\bar{R}ic=\lambda$. We obtain $\mathcal{A}(t,r)$ and $\mathcal{B}(t,r)$ as the following form,

\begin{equation}\label{Eq11}
\mathcal{B}(t,r)=\alpha(t)\left(a\lambda-\frac{a\Lambda}{3}r^{2}+\frac{b}{r}\right), ~~~ \mathcal{A}(t,r)=\left(\lambda-\frac{\Lambda}{3}r^{2}+\frac{b}{ar}\right)^{-1},
\end{equation}
where $a$ and $b$ are integral constant. The obtained solutions should approach
Newtonian gravity in the weak-field approximation. Thus, we have
\begin{equation}\label{Eq11.1}
  b\lambda=-2GM.
\end{equation}
Now, $\mathcal{B}(t,r)$ and $\mathcal{A}(t,r)$ in relativistic units $(c = G =1)$ are rewritten as follows
\begin{equation}\label{Eq11.2}
\mathcal{B}(t,r)=\alpha(t)\left(a\lambda-\frac{a\Lambda}{3}r^{2}-\frac{2M}{\lambda r}\right), ~~~ \mathcal{A}(t,r)=\left(\lambda-\frac{\Lambda}{3}r^{2}-\frac{2M}{a\lambda r}\right)^{-1}.
\end{equation}
As we know, the Schwarzschild space-time has the interior solution, which can infer the famous Tolman–Oppenheimer–Volkoff (TOV) equation \cite{Tolman,Oppenheimer}. The interior behavior of Finslerian space-time \eqref{Eq6} is worth exploring. Although, there are obstacles in constructing gravitational field equations in Finsler space-time. We will show that there is a self-consistent gravitational field equation in Finslerian space-time \eqref{Eq6}.
Using the expressions \eqref{Eq5}, \eqref{Eq5.1}, and \eqref{Eq10}, we obtain
\begin{equation}\label{Eq11.3}
\mathcal{S}=\frac{\mathcal{B}^{\prime\prime}}{\mathcal{A}\mathcal{B}}-\frac{1}{2}\frac{\mathcal{B}^{\prime}}{\mathcal{A}\mathcal{B}}
\bigg(\frac{\mathcal{A}^{\prime}}{\mathcal{A}}+\frac{\mathcal{B}^{\prime}}{\mathcal{B}}\bigg)-\frac{\ddot{\mathcal{A}}}{\mathcal{A}\mathcal{B}}
+\frac{1}{2}\frac{\dot{\mathcal{A}}}{\mathcal{A}\mathcal{B}}\bigg(\frac{\dot{\mathcal{A}}}{\mathcal{A}}+\frac{\dot{\mathcal{B}}}{\mathcal{B}}\bigg)
+\frac{2}{r\mathcal{A}}\bigg(\frac{\mathcal{B}^{\prime}}{\mathcal{B}}-\frac{\mathcal{A}^{\prime}}{\mathcal{A}}\bigg)-
\frac{2}{r^{2}}\bar{R}ic+\frac{2}{r^2\mathcal{A}}.
\end{equation}
From \eqref{Eq11.3}, we found that $\mathcal{S}$ depends on the radial
coordinate $r$ and $t$. So, it implies that $\mathcal{G}^{i}_{j}$ is direction-independent. Using the expression \eqref{Eq5.2}, we obtain
\begin{align}\label{Eq11.4}
  \mathcal{G}^{t}_{t}= & \frac{\mathcal{A}^{\prime}}{r\mathcal{A}^{2}}-\frac{1}{r^2\mathcal{A}}+\frac{\bar{R}ic}{r^2}, \\
  \mathcal{G}^{r}_{r}= & \frac{-\mathcal{B}^{\prime}}{r\mathcal{A}\mathcal{B}}-\frac{1}{r^2\mathcal{A}}+\frac{\bar{R}ic}{r^2}, \\
  \mathcal{G}^{t}_{r}= & \frac{\dot{\mathcal{A}}}{r\mathcal{A}\mathcal{B}}, \\
  \mathcal{G}^{\theta}_{\theta}=\mathcal{G}^{\phi}_{\phi}= & \frac{-1}{2r\mathcal{A}}\bigg(\frac{\mathcal{B}^{\prime}}{\mathcal{B}}-\frac{\mathcal{A}^{\prime}}{\mathcal{A}}\bigg)-
  \frac{\mathcal{B}^{\prime\prime}}{2\mathcal{A}\mathcal{B}}+
  \frac{1}{4}\frac{\mathcal{B}^{\prime}}{\mathcal{A}\mathcal{B}}
\bigg(\frac{\mathcal{A}^{\prime}}{\mathcal{A}}+\frac{\mathcal{B}^{\prime}}{\mathcal{B}}\bigg)+
\frac{\ddot{\mathcal{A}}}{2\mathcal{A}\mathcal{B}}-
\frac{1}{4}\frac{\dot{\mathcal{A}}}{\mathcal{A}\mathcal{B}}
\bigg(\frac{\dot{\mathcal{A}}}{\mathcal{A}}+\frac{\dot{\mathcal{B}}}{\mathcal{B}}\bigg).
\end{align}
We analyze the covariant conserve properties of the tensor $\mathcal{G}^{i}_{j}$. The covariant derivative of $\mathcal{G}^{i}_{j}$ in Finsler space-time is given by \cite{Bao}
\begin{equation}\label{Eq11.5}
\mathcal{G}^{i}_{j|i}=\frac{\delta}{\delta x^{i}}\mathcal{G}^{i}_{j}+\Gamma^{i}_{ik}\mathcal{G}^{k}_{j}-\Gamma^{k}_{ij}\mathcal{G}^{i}_{k},
\end{equation}
where $\frac{\delta}{\delta x^{i}}=\frac{\partial}{\partial x^{i}}-\frac{\partial\mathcal{G}^{k}}{\partial y^{i}}\frac{\partial}{\partial y^{k}}$ and $\Gamma^{i}_{ik}$ is the Chern connection. We obtain that $\mathcal{G}^{i}_{t|i}=\mathcal{G}^{i}_{r|i}=\mathcal{G}^{i}_{\theta|i}=\mathcal{G}^{i}_{\phi|i}=0$, if $\dot{\mathcal{A}}=0$. Assuming $Ric =-\Lambda$, and motivated
Einstein-type equation \cite{PStavrinos} in Finsler space-time, we add the cosmological constant in gravitational field equation \eqref{Eq5.3}. So we have
\begin{equation}\label{Eq11.6}
\mathcal{G}^{i}_{j}-\Lambda\delta^{i}_{j}=8\pi_{\mathcal{F}}G\mathcal{T}^{i}_{j},
\end{equation}
Equation \eqref{Eq11.6} is the gravitational field equation in Finslerian structure \eqref{Eq6}. Finslerian space-time properties \eqref{Eq6} are consistent with Einstein field equation \eqref{Eq11.6}.
The volume of $\bar{\mathcal{F}}$ in \eqref{Eq11.6} is indicated by $4\pi_{\mathcal{F}}$. If we regard $\mathcal{F}_{FS}$ \eqref{Eq5.6} instead of $\bar{\mathcal{F}}$, then $\pi_{\mathcal{F}}=\pi$. Let $p(r )$ and $\rho(r )$ be the pressure and energy density of the gravitational source, respectively.  Now, we regard the energy-momentum tensor as follows,
\begin{equation}\label{Eq11.7}
\mathcal{T}^{i}_{j}=diag\left(\rho(r),-p(r),-p(r),p(r)\right).
\end{equation}
The gravitational field equation \eqref{Eq11.6} converts to the following equations,
\begin{eqnarray}
  \frac{2p^{\prime}}{\rho+p} = -\frac{\mathcal{B}^{\prime}}{\mathcal{B}}, \label{Eq11.8}\\
  \frac{\lambda}{r^2}-\frac{1}{r^2\mathcal{A}}+\frac{\mathcal{A}^{\prime}}{r\mathcal{A}^2}-\Lambda = 8\pi_{\mathcal{F}}\rho, \label{Eq11.9}\\
   \Lambda-\frac{\lambda}{r^2}+\frac{1}{r^2\mathcal{A}}+\frac{\mathcal{B}^{\prime}}{r\mathcal{A}\mathcal{B}}= 8\pi_{\mathcal{F}}p.\label{Eq11.10}
\end{eqnarray}
By solving \eqref{Eq11.9}, we have
\begin{equation}\label{Eq11.11}
  \mathcal{A}^{-1}(t,r)=\lambda-\frac{2\mathfrak{M}(r)}{r}-\frac{1}{3}\Lambda r^{2},
\end{equation}
where $\mathfrak{M}(r)=\int_{0}^{r}4\pi_{\mathcal{F}}\rho(\mathfrak{r})\mathfrak{r}^2 d\mathfrak{r}$. Plugging \eqref{Eq11.10}
into \eqref{Eq11.8}, and using the solution \eqref{Eq11.11}, we have
\begin{equation}\label{Eq11.12}
  -p^{\prime}r^2=(\rho+p)\left(4\pi_{\mathcal{F}}pr^3+\mathfrak{M}-\frac{1}{3}\Lambda r^3\right)\left(\lambda-\frac{2\mathfrak{M}}{r}-\frac{1}{3}\Lambda r^{2}\right).
\end{equation}
If we put the Riemannian sphere \eqref{Eq5.5} instead of $\bar{\mathcal{F}}$ and $\Lambda=0$, then the expression \eqref{Eq11.12} becomes the well-known TOV equation.

The interior solution \eqref{Eq11.11} should be compatible with the exterior solution \eqref{Eq11} at the boundary of the gravitational source. Therefore, we get
\begin{equation}\label{Eq11.14}
a\lambda = 1.
\end{equation}
Last, combining the boundary condition \eqref{Eq11.14} with the requirement of Newtonian limit \eqref{Eq11.1}, we get the exterior solution $\mathcal{B}(t,r)$ and $\mathcal{A}(t,r)$ as
\begin{equation}\label{Eq11.13}
\mathcal{B}(t,r)=\alpha(t)\left(1-\frac{2M}{\lambda r}-\frac{1}{3\lambda}\Lambda r^{2}\right), ~~~
  \mathcal{A}(t,r)=\left(\lambda-\frac{2M}{r}-\frac{1}{3}\Lambda r^{2}\right)^{-1}.
\end{equation}
The function $\alpha(t)$ can be equal to 1 by defining a new time coordinate,
\begin{equation*}
t^{\prime}=\int^{t}\sqrt{\alpha(\mathfrak{t})}d\mathfrak{t}.
\end{equation*}
The Finsler structure \eqref{Eq6} is time-independent and identical to the static solution \cite{SK}. Therefore, the counterpart of Birkhoff’s theorem exists in Finslerian gravity.
This metric looks like a Schwarzschild (anti-)de Sitter with a deficit
solid angle. Because $\lambda=1-\varrho$ where $\varrho$  is the deficit solid angle parameter. Since the gravitational effect of the cloud from the string is the same as that caused by the angle of solid deficiency. If $\lambda\neq 1$, this solution also is similar to a BH surrounded by a cloud of strings by comparing with the results of Refs. \cite{Rodrigue,Toledo,Nekouee}.\\
The horizons are given by the zeros of $g_{00}$ i.e., the real roots of
\begin{equation}\label{Eq13}
r^{3}-\frac{3\lambda}{\Lambda}r+\frac{6M}{\Lambda}=0.
\end{equation}
The above equation has three real roots whenever $\Delta=\frac{q^2}{4}+\frac{p^{3}}{27}<0$ where $p=-\frac{3\lambda}{\Lambda}$ and $q=\frac{6M}{\Lambda}$,
\begin{equation}\label{Eq14}
\Delta=\frac{9M^{2}\Lambda-\lambda^{3}}{\Lambda^{3}}<0=\begin{cases}
\Lambda>0 ~~ \Rightarrow 9M^{2}\Lambda-\lambda^{3}<0 ~~ \Rightarrow ~~ \lambda>0,\\
\Lambda<0 ~~ \Rightarrow 9M^{2}\Lambda-\lambda^{3}>0 ~~ \Rightarrow ~~ \lambda<0.
\end{cases}
\end{equation}
For a solar mass, $2M \sim 10^{3}m$, on the other hand, $\Lambda=\frac{3}{a^{2}}\sim 10^{-60}m^{-2}$
i.e. $a^{2}\sim10^{60}m^{2}$, so, typically $2M \ll a$ i.e. $\frac{M}{a}\cong 10^{-27} \ll 1$, where $a$ is the scale factor. \eqref{Eq13} has three real roots, which are obtained as follows,

\begin{align}\label{Eq15}
r_{1}&=\frac{2}{\sqrt{3}}\sqrt{-p}\sin\bigg(\frac{1}{3}\sin^{-1}\bigg(\frac{3\sqrt{3}q}{2(\sqrt{-p})^{3}}\bigg)\bigg),\nonumber\\
r_{2}&=\frac{-2}{\sqrt{3}}\sqrt{-p}\sin\bigg(\frac{1}{3}\sin^{-1}\bigg(\frac{3\sqrt{3}q}{2(\sqrt{-p})^{3}}\bigg)+\frac{\pi}{3}\bigg),\nonumber\\
r_{3}&=\frac{2}{\sqrt{3}}\sqrt{-p}\cos\bigg(\frac{1}{3}\sin^{-1}\bigg(\frac{3\sqrt{3}q}{2(\sqrt{-p})^{3}}\bigg)+\frac{\pi}{6}\bigg),
\end{align}
$r_{1}$, $r_{3}$ are the positive roots, and $r_{2}$ is the negative root of the equation, which is not acceptable.
\begin{figure}[htpb]
\centering
\includegraphics[width=5in]{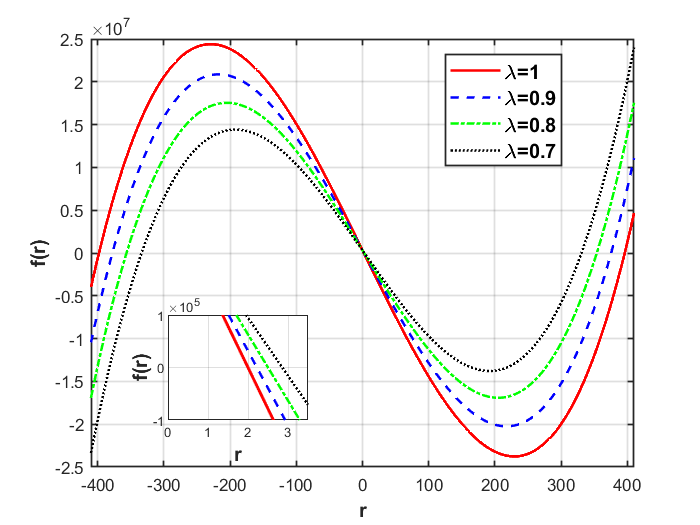}
\caption{\label{fig1} Horizons of the Finslerian Schwarzschild–de Sitter BH with $M=1$ and  $\Lambda=0.190371e-4.$}
\end{figure}

\begin{table}[htbp]
\centering
\begin{tabular}{|l|ccccc|}
\hline
Type & $\Lambda$ & $\lambda$ & $r_h$ & $r_c$ & $r_{ph}$\\
\hline
Schwarzschild & 0 & 1 & 2 & - & 3\\
Schwarzschild–de Sitter & 0.190371e-4 & 1 & 2.000050769 & 395.9685216 & 3\\
Finslerian SdS & 0.190371e-4 & 0.99 & 2.020254872 & 393.9684746 & 3.03030303\\
Finslerian SdS & 0.190371e-4 & 0.98 & 2.040871369 & 391.9581426 & 3.06122449\\
Finslerian SdS & 0.190371e-4 & 0.97 & 2.061913018 & 389.9373645 & 3.09278350\\
Finslerian SdS & 0.190371e-4 & 0.9  & 2.222299605 & 375.4849478 & 3.33333333\\
Finslerian SdS & 0.190371e-4 & 0.8  & 2.500123958 & 353.8061786 & 3.75\\
Finslerian SdS & 0.190371e-4 & 0.7  & 2.857354339 & 330.6929810 & 4.28571428\\
Finslerian SdS & 0.190371e-4 & 0.6  & 3.333725181 & 305.8130248 & 5\\
Finslerian SdS & 0.190371e-4 & 0.5  & 4.000812745 & 278.6800326 & 6\\
\hline
\end{tabular}
\caption{\label{tab:i} The values of event horizon ($r_h$), cosmological horizon ($r_c$), and photon radius ($r_{ph})$ for Schwarzschild, Schwarzschild–de Sitter, and Finslerian Schwarzschild–de Sitter BHs, (M=1).}
\end{table}
Table~\ref{tab:i} shows that as $\lambda$ decreases, the values of $r_{h}$ and $r_{ph}$ increase while $r_{c}$ decreases.
Also, for $\lambda<0.8$, the amount of $r_{ph}$ changes is very high.
\section{Geodesics of Finsler Space-time}\label{sec:3}
Triantafyllopoulos et al. \cite{Triantafyllopoulos} examined the space-like, null, and time-like trajectories on the Schwarzschild-Randers space-time.
Additionally, they numerically obtained the time-like trajectories and showed that a slight deviation exists with the geodesics of GR.
Therefore, they investigated the perturbation in the features of the classical Schwarzschild space-time.
Recently, the geodesics (time-like or null) on the Schwarzschild-Finsler–Randers space-time are investigated in \cite{Kapsabelis}, and the deviation from GR is demonstrated graphically. Additionally, the phase portraits and the effective potential of the model induced by them are analyzed and analogized with GR. Also, Kapsabelis et al. obtained the Schwarzschild-Finsler-Runder space-time deviation angle and showed that this model has a slight deviation from GR.
Cao \cite{Cao} has contemplated SR on the Finslerian space-time and examined that the space-like and the time-like modes could convert immediately into a light-like mode and vice-versa for the particle motion. But the time-like mode cannot smoothly convert into the space-like case.
Reddy \cite{Reddy} examined the motion of massive and massless particles in Schwarzschild Finslerian space-time and the scattering of particles near the Schwarzschild Finslerian BH. Additionally, indicated that if $\varepsilon=0$, the obtained results reduce to the Riemannian case.\\
In the present work, the Finslerian sphere donates to a perturbation of the classical Schwarzschild-de Sitter space-time properties and particle orbits. We will obtain Finslerian geodesics, which are generalizations of GR geodesics. If Finslerian parameter $\varepsilon=0$, we can retrieve the classical Schwarzschild-de Sitter space-time and thus the GR geodesics.\\
Randers-Finsler space \cite{Randers} has been classified by Bao et al. \cite{Bao1} with constant flag curvature. The following is a two-dimensional Randers-Finsler space with positive constant flag curvature $\lambda=1$,

\begin{equation}\label{Eq17}
\mathcal{\bar{F}}=\frac{\sqrt{\bigg(1-\varepsilon^{2}\sin^{2}(\theta)\bigg)y^{\theta}y^{\theta}+\sin^{2}(\theta)y^{\phi}y^{\phi}}}{1-\varepsilon^{2}\sin^{2}(\theta)}-
\frac{\varepsilon\sin^{2}(\theta)y^{\phi}}{1-\varepsilon^{2}\sin^{2}(\theta)},
\end{equation}
where $0\leq\varepsilon<1$.\\
The Lagrangian corresponding to \eqref{Eq6} assuming the Finslerian sphere takes the following form
\begin{equation}\label{Eq17.1}
\mathcal{L}=\frac{1}{2}\left(\alpha(t)B(r)y^{t}y^{t}-B(r)^{-1}y^{r}y^{r}-r^{2}\bar{\mathcal{F}}^{2}(\theta,\phi,y^{\theta},y^{\phi})\right).
\end{equation}
where $B(r)=1-\frac{\Lambda}{3}r^2-\frac{2M}{r}$. We assume the following generator of symmetry,
\begin{equation}
X=D_{1}\frac{\partial}{\partial t}+D_{2}\frac{\partial}{\partial r}+D_{3}\frac{\partial}{\partial \theta}+D_{4}\frac{\partial}{\partial \phi}+\dot{D}_{1}\frac{\partial}{\partial \dot{t}}+\dot{D}_{2}\frac{\partial}{\partial \dot{r}}+\dot{D}_{3}\frac{\partial}{\partial \dot{\theta}}+\dot{D}_{4}\frac{\partial}{\partial \dot{\phi}},
\end{equation}
where $D_{i}$ are functions of $t, r, \theta$ and $\phi$.
We apply the Noether symmetry theory to obtain  $\alpha(t)$. Now, we introduce the following
equations,
\begin{equation}
\dot{D_{i}}=\frac{\partial D_{i}}{\partial t}\dot{t}+\frac{\partial D_{i}}{\partial r}\dot{r}+\frac{\partial D_{i}}{\partial \theta}\dot{\theta}+\frac{\partial D_{i}}{\partial \phi}\dot{\phi}.
\end{equation}
According to Noether symmetry and Lie condition as $L_{X}\mathcal{L}=0$, we obtain a system of coupled partial differential equations as follows\\
\begin{align*}
0&=D_{1}\dot{\alpha}(t)B(r)+D_{2}\alpha(t)B^{\prime}(r)+2\alpha(t)B(r)\frac{\partial D_{1}}{\partial t},\\
0&=D_{2}\frac{B^{\prime}(r)}{B^{2}(r)}-\frac{2}{B(r)}\frac{\partial D_{2}}{\partial r},\\
0&=2\alpha(t)B(r)\frac{\partial D_{1}}{\partial r}-\frac{2}{B(r)}\frac{\partial D_{2}}{\partial t},\\
0&=2\alpha(t)B(r)\frac{\partial D_{1}}{\partial \theta}-\frac{2r^{2}}{1-\epsilon^2\sin^2\theta}\frac{\partial D_{3}}{\partial t},\\
0&=2\alpha(t)B(r)\frac{\partial D_{1}}{\partial \phi}-\frac{2r^{2}\sin^2\theta(1+\epsilon^2\sin^2\theta)}{(1-\epsilon^2\sin^2\theta)^2}\frac{\partial D_{4}}{\partial t},\\
0&=-\frac{2}{B(r)}\frac{\partial D_{2}}{\partial \theta}-\frac{2r^{2}}{1-\epsilon^2\sin^2\theta}\frac{\partial D_{3}}{\partial r},\\
0&=-\frac{2}{B(r)}\frac{\partial D_{2}}{\partial \phi}-\frac{2r^{2}\sin^2\theta(1+\epsilon^2\sin^2\theta)}{(1-\epsilon^2\sin^2\theta)^2}\frac{\partial D_{4}}{\partial r}, \\
0&=\frac{2r^2\epsilon^3\sin^3\theta\cos\theta}{(1-\epsilon^2\sin^2\theta)^2}D_{3}+\frac{2r^2\epsilon\sin^2\theta}{1-\epsilon^2\sin^2\theta}\frac{\partial D_{3}}{\partial \theta},\\
0&=-\frac{2r}{1-\epsilon^2\sin^2\theta} D_{2}-\frac{2r^{2}\epsilon^2\sin\theta\cos\theta}{(1-\epsilon^2\sin^2\theta)^2}D_{3}-\frac{2r^2}{1-\epsilon^2\sin^2\theta}\frac{\partial D_{3}}{\partial \theta},\\
0&=-\frac{2r\sin^2\theta(1+\epsilon^2\sin^2\theta)}{(1-\epsilon^2\sin^2\theta)^2} D_{2}-\frac{2r^{2}\sin\theta\cos\theta(1+3\epsilon^2\sin^2\theta)}{(1-\epsilon^2\sin^2\theta)^3}D_{3}\\\nonumber
&~~~-\frac{2r^2\sin^2\theta(1+\epsilon^2\sin^2\theta)}{(1-\epsilon^2\sin^2\theta)^2}\frac{\partial D_{4}}{\partial \phi},\\
0&=\frac{4r\epsilon\sin^2\theta}{(1-\epsilon^2\sin^2\theta)^2}D_{2}+\frac{2r^2\epsilon\sin^2\theta}{(1-\epsilon^2\sin^2\theta)^2}\frac{\partial D_{4}}{\partial\phi}+\frac{4r^2\epsilon\sin\theta\cos\theta}{(1-\epsilon^2\sin^2\theta)^3}D_{3},\\
0&=\frac{2r^2\epsilon\sin^3\theta\cos\theta(1-\epsilon^2\sin^2\theta)}{(1-\epsilon^2\sin^2\theta)^3}D_{3}+\frac{2r^2\epsilon\sin^4\theta}{(1-\epsilon^2\sin^2\theta)^2}\frac{\partial D_{4}}{\partial \phi},
\end{align*}
\begin{align*}
0&=-\frac{2r^2}{1-\epsilon^2\sin^2\theta} \frac{\partial D_{3}}{\partial \phi}-\frac{2r^{2}\sin^2\theta(1+\epsilon^2\sin^2\theta)}{(1-\epsilon^2\sin^2\theta)^2}\frac{\partial D_{4}}{\partial \theta},\\
0&=\frac{2r^2\epsilon\sin^2\theta}{1-\epsilon^2\sin^2\theta} \frac{\partial D_{3}}{\partial \phi}+\frac{2r^{2}\epsilon\sin^4\theta}{(1-\epsilon^2\sin^2\theta)^2}\frac{\partial D_{4}}{\partial \theta},\\
0&=\frac{\partial D_{4}}{\partial t}=\frac{\partial D_{4}}{\partial r}=\frac{\partial D_{4}}{\partial \theta}=\frac{\partial D_{3}}{\partial t}=\frac{\partial D_{3}}{\partial r}.
\end{align*}
The solution of the system is $D_{4}=$constant, $D_{2}=D_{3}=0$, $\alpha(t)=\frac{c}{D_{1}^2(t)}$, and $D_{1}(t)\neq 0$. Noether's theorem shows that $\alpha(t)$ is an arbitrary function of $t$, so as in the previous section, we can consider it equal to 1 without losing the generality of the problem.\\
The conserved values along geodesics in the frame $\{t, r,\theta,\phi\}$ with metric \eqref{Eq6} are energy $E$ and angular momentum $L$.
In order to find $V_{eff}$ for this BH, we can present the following strategy. One can define the conjugate momentum $P_{\mu}$ to the coordinate $x^{\mu}$ as
\begin{equation}\label{Eq18}
P_{\mu}\equiv\frac{\partial \mathcal{L}}{\partial\dot{x}^{\mu}}=g_{\mu\nu}\dot{x}^{\nu},
\end{equation}
where
\begin{equation}\label{Eq19}
\dot{x}^{\mu}\equiv\frac{dx^{\mu}}{d\tau}=u^{\mu}.
\end{equation}
In terms of the conjugate momenta, the Euler-Lagrange equations can be written as
\begin{equation}\label{eq20}
 \frac{d}{d\tau}P_{\mu}=\frac{\partial\mathcal{L}}{\partial x^{\mu}}.
\end{equation}
The mentioned BH is stationary and axisymmetric, so there are two Killing vectors $k^{\mu}=(1,0,0,0)$ and $m^{\mu}=(0,0,0,1)$
which are time-like and space-like, respectively.
To obtain the conserved constants, one can find the geodesics equations by the above Killing vectors. Then, for constants, we can find $E=-\partial_t. U$ and $L=\partial_\phi. U$
where $U$, $E$, and $L$ correspond to four-velocity and its energy and angular momentum in our system. In this work, we restrict ourselves to the equatorial plane $(\theta =\frac{\pi}{2})$ implying
$U^\theta=0 $. The radial geodesic is obtained by solving the equation
$U.U=-1.$ Therefore, the canonical momenta $P_t$ and $P_{\phi}$ will be conserved, which $E$ and $L$ are interpreted as energy and angular moment per unit mass, respectively as
\begin{equation}\label{Eq21}
E\equiv -k_{\mu}u^{\mu}=-g_{t\mu}u^{\mu}=-P_{t},
\end{equation}
\begin{equation}\label{Eq22}
L\equiv m_{\mu}u^{\mu}=-g_{\varphi\mu}u^{\mu}=P_{\phi},
\end{equation}
where $E$ is the energy at infinity and $L$ is the angular momentum. In order to obtain $\dot{t}=\frac{dt}{d\tau}$ and $\dot{\phi}=\frac{d\phi}{d\tau}$, we have to consider \eqref{Eq21} and \eqref{Eq22}, so we obtain the following equation,
\begin{equation}\label{Eq23}
\dot{t}=\frac{-E}{B(r)},
\end{equation}
and
\begin{equation}\label{Eq24}
\dot{\phi}=\frac{L(1+\varepsilon^{2})}{r^{2}}=u^{\phi}.
\end{equation}
Furthermore, the subject of effective potential and center mass energy plays a crucial role in the collision of two particles on the BH background. As mentioned in section~\ref{sec2}, the Finslerian structure along the geodesic is constant. The geodesic equations are expressed with this condition as follows,
\begin{equation}\label{Eq25}
g_{\mu\nu}u^{\mu}u^{\nu}=\mathcal{F}^2,
\end{equation}
where $\mathcal{F}=0$ for light photons and $\mathcal{F}=1$ for massive particles.
According to the \eqref{Eq25}, we have
\begin{equation}\label{Eq26}
B(r)\dot{t}^{2}-\frac{1}{B(r)}\dot{r}^{2}-\frac{r^{2}}{(1+\varepsilon)^{2}}\dot{\phi}^{2}=\mathcal{F}^2.
\end{equation}

\subsection{Light Photon around Black Hole}\label{subsec:31}

The gravitational lensing problem was first studied by Einstein and Eddington \cite{Eddington} in GR, but Darwin solved it \cite{Darwin,Darwin2}. For this purpose, he obtained the null geodesic around the photon sphere of the Schwarzschild black hole in Minkowski's flat space-time. Using the same commonly used method \cite{Claudel,Virbhadra1,Virbhadra2,Virbhadra3,Virbhadra4,Virbhadra5}, we consider null geodesics in the Finslerian Schwarzschild-de Sitter system.

Now we assume that $\mathcal{F}=0$ and using \eqref{Eq26} we obtain,
\begin{equation}\label{Eq27}
\frac{E^{2}}{B(r)}-\frac{\dot{r}^{2}}{B(r)}-\frac{L^{2}(1+\varepsilon)^{2}}{r^{2}}=0,
\end{equation}
by simplifying the above equation we have
\begin{equation}\label{Eq28}
\dot{r}^{2}=E^{2}-B(r)\bigg(\frac{L^{2}(1+\varepsilon)^{2}}{r^{2}}\bigg),
\end{equation}
where
\begin{equation}\label{Eq29}
V_{eff}=\bigg(1-\frac{\Lambda}{3}r^{2}-\frac{2M}{r}\bigg)\bigg(\frac{L^{2}(1+\varepsilon)^{2}}{r^{2}}\bigg).
\end{equation}
\begin{figure}[htpb]
\centering
\includegraphics[width=5in]{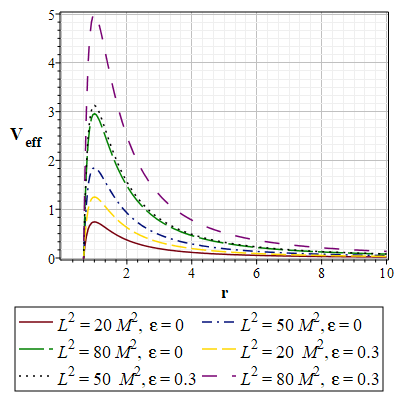}
\caption{\label{fig2} The $V_{eff}$ in Riemannian and Finsler modes.}
\end{figure}
We show the effective potentials of the Finsler and Riemannian models by comparing the three for $ L^{2}=20M^{2}, 50M^{2}$, and $80M^{2}$. As we can see in figure \ref{fig2}, the difference between Riemannian and Finsler cases when the angular momentum becomes weak, the relevant difference is rather small.\\

Differentiation of \eqref{Eq28} with respect to the affine parameter
$\tau$ allows us to find a Newton-type law of effective force
for the radial coordinate given by
\begin{equation}\label{Eq30}
\ddot{r}=-\frac{dV_{eff}}{dr}=-\frac{2 L^{2}(1+\varepsilon)^{2}}{r^{3}}+\frac{6ML^{2}(1+\varepsilon)^{2}}{r^{4}}.
\end{equation}
This radial acceleration is a sign of the radial coordinate's change as a result of the photon trajectory's curvature.
The above calculation shows that this acceleration is zero for radial photons with $L = 0$. It is important to note that this calculation does not depend on the cosmological constant $\Lambda$, which suggests that the highest effective potential is located where $r _{ph} = 3M$.
\begin{eqnarray}\label{Eq31}
r \rightarrow r(\phi), ~~ \dot{r}\rightarrow \frac{dr}{d\phi}\dot{\phi}=\frac{dr}{d\phi}\frac{L}{r^{2}(\phi)(1+\varepsilon)^{2}}.
\end{eqnarray}
After a bit of calculation, the shapes of the photon geodesics are given by the functions $r(\phi)$, which satisfy \eqref{Eq26} with $\mathfrak{F}=0$ that now reads,
\begin{equation}\label{Eq32}
\bigg(\frac{dr}{d\phi}\bigg)^{2}-r^{4}(\phi)\frac{E^{2}}{L^{2}}+r^{2}(\phi)f(r)\bigg((1+\varepsilon)^{2}\bigg)=0.
\end{equation}
Circular geodesics on the photon sphere and corresponding spiral geodesics are two different varieties of solutions to this equation \cite{Darwin}. The circular geodesics simultaneously satisfy  the conditions
$\frac{dr(\phi)}{d\phi}=\frac{d^{2}r(\phi)}{d\phi^{2}}=0$, giving the radius of the photon sphere $r_{ph} = 3M$ and the
mandatory condition
\begin{equation}\label{Eq33}
L_{ph}=\pm\frac{3\sqrt{3}ME_{ph}}{\sqrt{1-9\Lambda M^{2}}(1+\varepsilon)},
\end{equation}
derived in ref. \cite{Perlick}. Thus, the photons with circular geodesics
are trapped on the photon sphere without escaping outside.
Furthermore, by substituting the condition \eqref{Eq33} in \eqref{Eq32}
we obtain the equation
\begin{equation}\label{Eq34}
27M^{2}\bigg(\frac{dr}{d\phi}\bigg)^{2}-54M^{3}(1+\varepsilon)^{2}r+27M^{2}(1+\varepsilon)^{2}r^{2}-(1+\varepsilon)^{2}r^{4}=0,
\end{equation}
which is independent of $\Lambda$.
This equation permits the solutions in addition to the circular geodesics.
\begin{equation}\label{Eq35}
r_{(\pm)}(\phi)=3M\frac{(A^{\pm}+6M)^{2}}{(6M+A^{\pm})^{2}-36MA^{\pm}},
\end{equation}
where $A=\exp((1+\varepsilon)\phi)$, $A^{-}=\frac{1}{A}$, and $A^{+}=A$.\\
\begin{figure}[htpb]
\centering
\mbox{{\includegraphics[width=2.9in]{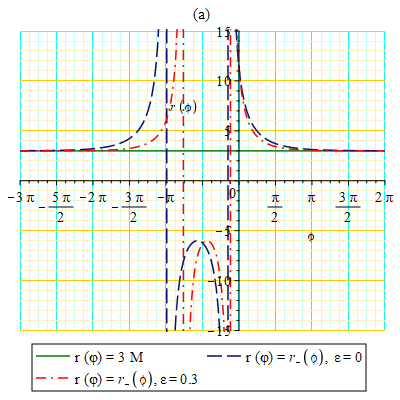}}
{\includegraphics[width=2.9in]{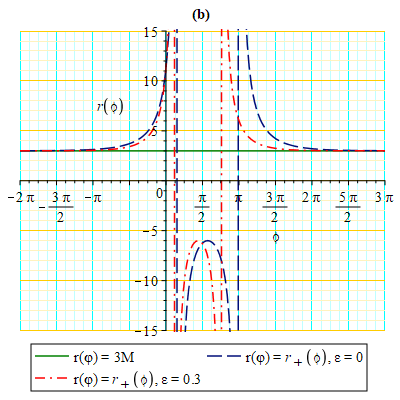} }}
\caption{\label{fig3} The behavior of functions $r_{(\pm)}(\phi)$ of the spiral photon geodesics.}
\end{figure}
We find that the functions $r_{(\pm)}(\phi)$ are defined on the domains $(-\infty,\phi_{\pm_{1}})\cup(\phi_{\pm_{2}},+\infty)$, where $\phi_{+_{1,2}}=\frac{\ln(6M(2\pm\sqrt{3}))}{1+\varepsilon}$ and $\phi_{-_{1,2}}=-\frac{\ln(6M(2\pm\sqrt{3}))}{1+\varepsilon}$, as their values are negative between the vertical asymptotes and have no physical significance. Intriguingly, the BH mass does not affect this distance.
\begin{equation*}
\lvert \phi_{(\pm)2}-\phi_{(\pm)1}\lvert\simeq\frac{2.6339}{(1+\varepsilon)}.
\end{equation*}
In figure \ref{fig3}, it can be seen that this distance is smaller in the Finsler state than in the Riemann state.
In the physical domain, these trajectories remain outside the
photon sphere, $r_{(\pm)}(\phi) > 3M$, but approaching to this for
large $|\phi|$ since
\begin{equation}\label{Eq36}
\lim _{\phi \rightarrow \pm \infty}r_{(\pm)}(\phi)=3M.
\end{equation}

\begin{figure}[htpb]
\centering
\mbox{{\includegraphics[width=2.9in]{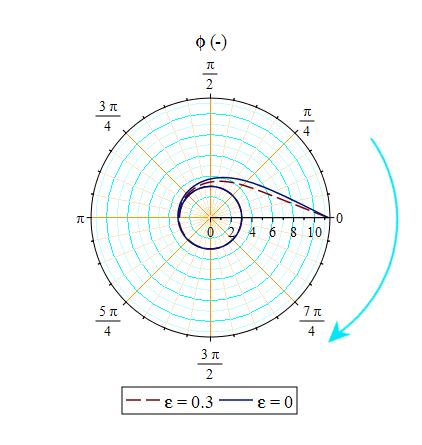}}
{\includegraphics[width=2.9in]{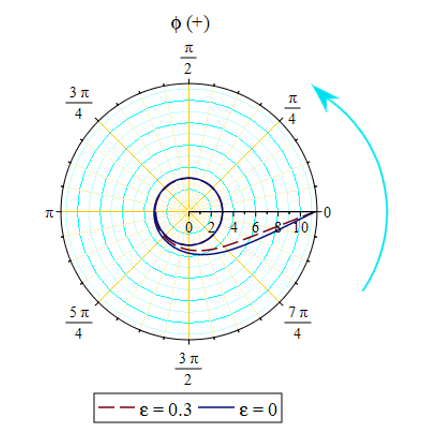} }}
\caption{\label{fig4} The behavior of functions $r_{(\pm)}(\phi)$ of the spiral photon geodesics in polar coordinate.}
\end{figure}
Figure \ref{fig4} illustrates the spiral geodesics spined out near the photon sphere escaping outside just when $\phi$ is close to the values $\phi_{+_{1,2}}=\frac{\ln(6M(2\pm\sqrt{3}))}{1+\varepsilon}$ and $\phi_{-_{1,2}}=-\frac{\ln(6M(2\pm\sqrt{3}))}{1+\varepsilon}$. In the Finslerian case, the curve contracts inwards, although these changes are slight.\\

\subsection{The Effective Potential under Circular Orbit Conditions}\label{subsec:32}
Here, we assume that $\mathcal{F}=1$. Using outcomes \eqref{Eq21} and \eqref{Eq22}, as well as the time-like geodesic provided by \eqref{Eq26}, we obtain,
\begin{equation}\label{Eq37}
\frac{1}{2}\dot{r}^{2}+\frac{1}{2}\bigg(\frac{L^{2}(1+\varepsilon)^2}{r^{2}}-\frac{2ML^{2}(1+\varepsilon)^2}{r^{3}}-\frac{2M}{r}-
\frac{\Lambda}{3}r^{2}\bigg)=\underbrace{\frac{1}{2}\bigg(E^2-1+\frac{\Lambda L^{2}(1+\varepsilon)^2}{3}\bigg)}_\text{C},
\end{equation}
where $C$ is a constant based on the initial circumstances of the motion. So, we can define the effective potential as follows,
\begin{equation}\label{Eq38}
V_{eff}=\frac{1}{2}\bigg(-\frac{2M}{r}-
\frac{\Lambda}{3}r^{2}+\frac{L^{2}(1+\varepsilon)^2}{r^{2}}-\frac{2ML^{2}(1+\varepsilon)^2}{r^{3}}\bigg).
\end{equation}
Now, we can obtain the critical points from the effective potential obtained from \eqref{Eq38},
For this purpose, we initiate the analysis with the case where the Cosmological constant is zero.
 This case is significant because the condition $\Lambda\neq 0$ only affects local physics on astrophysical scales or larger. Thus, for example, no results on the scale of the Solar System will be affected with the condition $\Lambda\neq 0$.\\
 For $\Lambda=0$, the effective potential is given by the following form,
\begin{equation}\label{Eq38.1}
V_{eff}\mid_{\Lambda=0}=\frac{1}{2}\bigg(-\frac{2M}{r}+\frac{L^{2}(1+\varepsilon)^2}{r^{2}}-\frac{2ML^{2}(1+\varepsilon)^2}{r^{3}}\bigg).
\end{equation}
With the derivative of this equation with respect to r, we have
\begin{equation}\label{Eq38.2}
\frac{dV_{eff}\mid_{\Lambda=0}}{dr}=\frac{M}{r^2}-\frac{L^{2}(1+\varepsilon)^2}{r^{3}}+\frac{3ML^{2}(1+\varepsilon)^2}{r^{4}}.
\end{equation}
by solving \eqref{Eq38.2}, If $L\rightarrow\infty$, we obtain
\begin{equation}\label{Eq38.3}
r_{c}\approx \frac{L^2(1+\varepsilon)^2\pm L^2(1+\varepsilon)^2(1-\frac{6M^2}{L^2(1+\varepsilon)^2})}{M}=\begin{cases}
r_{c1}=\frac{L^2(1+\varepsilon)^2}{M},\\
r_{c2}=3M,
\end{cases}
\end{equation}
where $r_{c}$ is the distance at which the orbits are circular. The first root corresponds to the stable equilibrium \cite{Carroll}. The second root is proportional to the Schwarzschild radius and corresponds to an unstable equilibrium.
When the discriminant of quadratic equation \eqref{Eq38.2} is zero, the minimum angular momentum to obtain bound orbits is as follows,
\begin{equation}\label{Eq38.4}
L_{min}=\frac{2\sqrt{3}M}{(1+\varepsilon)}.
\end{equation}
If the angular momentum takes this value, the two distances obtained for the circular orbits from quadratic equation \eqref{Eq38.2} become the same, and they satisfy the condition of the saddle point, so we have
\begin{equation}\label{Eq38.5}
r_{cs}=6M.
\end{equation}
Since this result corresponds to the saddle point condition, we can find an equivalent solution. For this purpose, we must solve simultaneously the equations obtained from $\frac{dV_{eff}\mid_{\Lambda=0}}{dr}=\frac{d^{2}V_{eff}\mid_{\Lambda=0}}{dr^2}=0$.
If we substitute the second root obtained from \eqref{Eq38.3} inside \eqref{Eq38.2}, we have
\begin{equation}\label{Eq38.6}
\frac{dV_{eff}\mid_{\Lambda=0}}{dr}=\frac{M}{9M^2}-\frac{L^{2}(1+\varepsilon)^2}{27M^3}+\frac{3ML^{2}(1+\varepsilon)^2}{81M^4}=\frac{1}{9M}.
\end{equation}
So scale $r_{c2}=3M$ is not the exact solution \eqref{Eq38.2}. This scale becomes an exact solution if the approximation $L \gg 2M$ is satisfied. In such a case, the corresponding expression to Newtonian contribution $\frac{M}{r^2}\approx0$ at short scales r. Using this condition, for short scales $(r\rightarrow 2M)$, The effective potential can be approximately in the following form,
\begin{equation}\label{Eq38.7}
V_{eff}\mid_{\Lambda=0}\approx\frac{L^{2}(1+\varepsilon)^2}{2r^{2}}-\frac{ML^{2}(1+\varepsilon)^2}{r^{3}},
\end{equation}
by deriving the above equation with respect to r, we have
\begin{equation}\label{Eq38.8}
\frac{dV_{eff}\mid_{\Lambda=0}}{dr}\approx-\frac{L^{2}(1+\varepsilon)^2}{r^{3}}+\frac{3ML^{2}(1+\varepsilon)^2}{r^{4}},
\end{equation}
By solving \eqref{Eq38.8}, we obtain
\begin{equation}\label{Eq38.9}
r_{1}=3M,
\end{equation}
which corresponds to the second root obtained in \eqref{Eq38.3}.
As in the previous case, $r_{c1}=\frac{L^{2}(1+\varepsilon)^2}{M}$ is not the exact solution of the equation $\frac{dV_{eff}\mid_{\Lambda=0}}{dr}$. This scale is the exact solution of the equation if the term GR contribution is negligible ($\frac{3ML^{2}(1+\varepsilon)^2}{r^4}\approx 0$). Then, for large scales, under the presumption $L \gg 2M$, we have
\begin{equation}\label{Eq38.10}
V_{eff}\mid_{\Lambda=0}\approx\frac{-M}{r}+\frac{L^{2}(1+\varepsilon)^2}{2r^{2}},
\end{equation}
by ignoring the GR contribution, we can calculate the extremal conditions for the potential \eqref{Eq38.10} and then obtain
\begin{equation}\label{Eq38.11}
\frac{dV_{eff}\mid_{\Lambda=0}}{dr}\approx\frac{M}{r^2}-\frac{L^{2}(1+\varepsilon)^2}{r^{3}}=0 ~~ \Rightarrow ~~ r_{2}=\frac{L^{2}(1+\varepsilon)^2}{M}.
\end{equation}

\begin{figure}[htpb]
\centering
\includegraphics[width=6in]{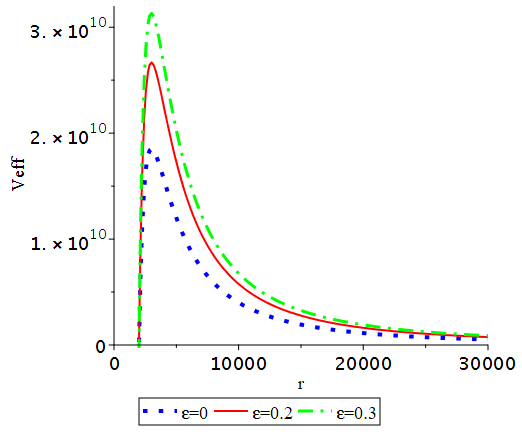}
\caption{\label{fig5} The effective potential in both Riemannian and Finslerian cases. Here $\mathbb{B}=10$ and $L = 10^{9}$.}
\end{figure}

\begin{figure}[htpb]
\centering
\includegraphics[width=6in]{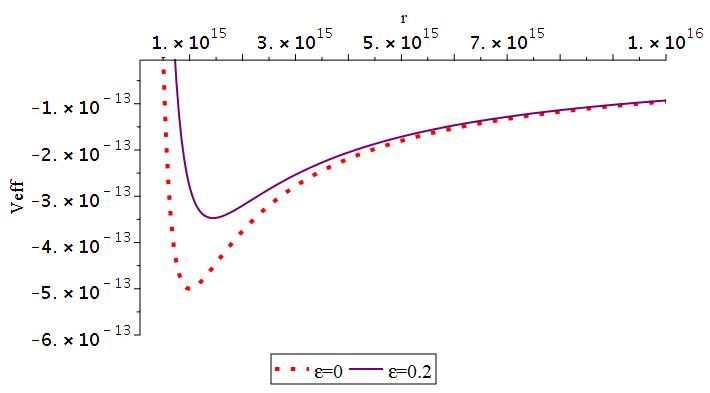}
\caption{\label{fig6} Riemannian and Finslerin valid effective potential for scale $2M \ll r \ll (\frac{3M}{\Lambda})^{\frac{1}{3}}$ and $L \gg 2M$ for $\mathbb{B}=10, L=10^{9}$. The shape of the figure will not change for different values of $\mathbb{B}$.}
\end{figure}

\begin{figure}[htpb]
\centering
\includegraphics[width=6in]{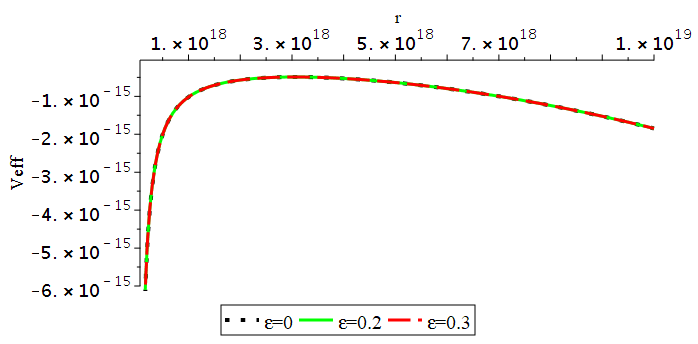}
\caption{\label{fig7} Riemannian and Finslerin valid effective potential for scale $ r \approx(\frac{3M}{\Lambda})^{\frac{1}{3}}$ and $L \gg 2M$ for special value of $\mathbb{B}$.}
\end{figure}

We can repeat the previous method for $\Lambda\neq0$. The Newtonian gravitational potential is the first term on the right-hand side of \eqref{Eq38}, followed by the $\Lambda$ contribution, which reproduces a repulsive effect, and the centrifugal force, which has the same shape in both Newtonian gravity and GR. The GR correction to the effective potential is the final term. This final contribution is significant at small scales, that is, comparable to the gravitational radius 2GM. Normally, $\frac{1}{\sqrt{\Lambda}}\gg 2GM$, and if $\frac{1}{\sqrt{\Lambda}}\approx 2GM$, the BH is then close to reaching its maximal mass before it becomes a naked singularity, here we assume $G=1$. At large scales, with $L \gg 2M$, the previous potential can be approximated to
\begin{equation}\label{Eq39}
V_{eff}\approx \frac{-M}{r}-\frac{\Lambda}{6}r^{2}+\frac{L^{2}(1+\varepsilon)^2}{2r^{2}},
\end{equation}
where $\frac{-ML^{2}(1+\varepsilon)^{2}}{r^{3}}$, the GR correction term, has been disregarded. The following equation results from $\frac{dV_{eff}(r)}{dr}=0$,
\begin{equation}\label{Eq40}
r^{4}-\frac{3M}{\Lambda}r+\frac{3L^{2}(1+\varepsilon)^2}{\Lambda}=0.
\end{equation}
This is a reduced fourth-order polynomial as has been defined in the following form,
\begin{equation}
r^{4} + \mathcal{P}r^{2} + \mathcal{Q}r + \mathcal{K} = 0.
\end{equation}\label{Eq41}
We can evaluate using the typical reduced form recommended by \eqref{Eq41}. Following a standard technique ($y^{3}+2\mathcal{P}y^{2}+(\mathcal{P}^{2}-4\mathcal{K})y-\mathcal{Q}^{2}=0$), the discriminant for the quartic polynomial \eqref{Eq40} can be found as follows,
\begin{equation}\label{Eq42}
y^{3}\underbrace{-\frac{12L^{2}(1+\varepsilon)^2}{\Lambda}}_\text{p}y\underbrace{-\frac{9M^{2}}{\Lambda^{2}}}_\text{q}=0.
\end{equation}
According to the solutions of the related third-order \eqref{Eq42},
solutions of the reduced \eqref{Eq40} are given by
\begin{align}
r_{1}&=\frac{1}{2}\bigg(\sqrt{y_{1}}+\sqrt{y_{2}}-\sqrt{y_{3}}\bigg),\\\nonumber
r_{2}&=\frac{1}{2}\bigg(\sqrt{y_{1}}-\sqrt{y_{2}}+\sqrt{y_{3}}\bigg),\\\nonumber
r_{3}&=\frac{1}{2}\bigg(-\sqrt{y_{1}}+\sqrt{y_{2}}+\sqrt{y_{3}}\bigg),\\\nonumber
r_{4}&=\frac{1}{2}\bigg(-\sqrt{y_{1}}-\sqrt{y_{2}}-\sqrt{y_{3}}\bigg).
\end{align}
In a quadratic polynomial equation, this discriminant functions similarly to the root square term. The sign of the discriminant determines the type of solution that is found. It can be proved that there are no actual solutions if $\triangle<0$. We can obtain actual solutions if $\triangle > 0$. Then $\triangle = 0$ denotes a limit where we obtain the maximum angular momentum allowing for still constrained orbits. The formula for the greatest angular momentum is
\begin{equation}\label{Eq42.1}
L_{max}=\frac{3^{\frac{2}{3}}}{2(1+\varepsilon)}\bigg(\frac{M^{2}}{2\sqrt{\Lambda}}\bigg)^{\frac{1}{3}},
\end{equation}
where $\triangle=(\frac{p}{3})^{3}+(\frac{q}{2})^{2}$.\\
By assuming $\triangle > 0$, the quartic equation \eqref{Eq40} has the following solutions
\begin{equation}\label{Eq43}
r^{\star}_{1}=\frac{L^{2}(1+\varepsilon)^2}{M}, ~~~ r^{\star}_{2}=4^{\frac{1}{3}}\sqrt{\frac{L_{max}(1+\varepsilon)}{\sqrt{\Lambda}}}-\frac{1}{2\mathbb{B}^{2}}\sqrt{\frac{L_{max}(1+\varepsilon)}{\sqrt{\Lambda}}},
\end{equation}
where $\mathbb{B}=\frac{L_{max}}{L}$. $r^{\star}_{1}$ corresponds to $r_{c1}$. The second solution is new because it depends on $\Lambda$ that was previously removed. If $L = L_{max}$, we have $\mathbb{B}= 1$. The result corresponds to the condition of the second saddle point different from the one previously defined for the result \eqref{Eq38.4}. This new saddle point on large scales is defined as follows
\begin{equation}\label{Eq44}
r^{\star}_{1}=r^{\star}_{2}=r_{s}=\sqrt{\frac{L_{max}(1+\varepsilon)}{\sqrt{\Lambda}}}.
\end{equation}
By substituting \eqref{Eq42.1}, we obtain
\begin{equation}\label{Eq45}
r^{\star}_{1}=r^{\star}_{2}=r_{s}=\frac{3^{\frac{1}{3}}}{2}\bigg(\frac{2M}{\Lambda}\bigg)^{\frac{1}{3}}.
\end{equation}
Figure \ref{fig5} displays the effective potential curve for scales where the gravitational radius GM is the dominant one and corresponds to $r_{1}=3M$.
Figure \ref{fig6} displays the effective potential curve under the condition $L\gg2M$ at large scales. This part of the curve conforms to the stable orbit condition. This part of the effective potential will not change when the cosmological constant effects are included or disappear.
The result \eqref{Eq45} corresponds to the local maximum shown in figure \ref{fig7}.
\section{Conclusion}\label{sec:4}
The present work expands the classical Schwarzschild-de Sitter solution in the Finslerian framework.
Therefore, it presents a deviation in the geometric structure and, thus, in the theory of gravity in the Riemannian framework.
Manjunatha et al. \cite{SK} investigated a static solution of Finslerian Schwarzschild-de Sitter.
But, we analyzed time-dependent Finsler space-time.
We have generalized Einstein field equations with $\Lambda\neq0$ in the Finslerian framework and thus found a Finslerian Schwarzschild–de Sitter solution with the symmetry of Finslerian sphere from the Finsler structure \eqref{Eq6}.
The suggested Einstein field equation \eqref{Eq11.6} is consistent with the features of the Finsler structure \eqref{Eq6}.
If the energy-momentum tensor is zero, Einstein field equation \eqref{Eq11.6} reduces to the vacuum field equation. We understand
that the $Ric=-\Lambda$ has Schwarzschild–de Sitter metric as its solution.
If the Finslerian parameter $\varepsilon=0$, our obtained solution corresponds to the Schwarzschild–de Sitter solution.
In GR, the Birkhoff theorem ensures that the solution of the vacuum field equation with spherical symmetry should be static.
It implies that its exterior solution should be Schwarzschild metric regardless of the evolution of the gravitational source.
It is essential to examine such a subject in Finsler space-time.
Defining a new time coordinate, we showed that the Finsler structure \eqref{Eq11.6} is exhaustively time-independent and equal to the static solution.
On the other hand, with the help of the Noether theorem, we obtained the value of $\alpha(t)$ as an arbitrary function of $t$.
Without losing the generality of the problem, $\alpha(t)$ can be made to equal 1.\\
Since the gravitational effect of a cloud of strings is the same as that caused by a solid deficiency angle.
If $\lambda\neq1$, this solution is similar to a BH surrounded by a cloud of strings.
Here, we can see that the role of curvature and matter is the same. Instead of adding matter,  the same results are achieved by changing the geometry, and Finsler geometry is a suitable option for changing the geometry.\\
Using \eqref{Eq1}, it is proved that the Finslerian metric is constant along the geodetic \cite{SK}.
So, we investigated the geodesic analytical form of the Finslerian space-time with the symmetry of the Finslerian sphere based on particle energy and angular momentum along the geodesic (null or time-like) Finsler space-time.
Using the null geodesics, we obtained the geodesics $r(\pm)(\varphi)$.
They are the closest paths to the photon sphere of the first photons that can be observed at the boundary of the black hole shadow. Therefore, these photons are essential for analyzing shadow and the associated redshift. We found that there is a slight deviation from the Riemannian model in $r_{\pm}(\phi)$ and effective potential.\\
We obtained the critical points for the effective potential derived from \eqref{Eq38} for two cases $\Lambda=0$ and $\Lambda\neq 0$.
Figure \ref{fig5} shows the effective potential curve for scales where the gravitational radius GM is the dominant one and corresponds to $r_{1}=3M$.
Although the critical point does not depend on the parameter $\varepsilon$, we have shown that the effective potentials in the Finslerian and Riemannian models do not coincide, and this difference increases with the increase of the Finslerian parameter.\\
Figure \ref{fig6} shows the effective potential curve under the condition $L\gg2M$ at scale $2M \ll r \ll (\frac{3M}{\Lambda})^{\frac{1}{3}}$. This part of the curve conforms to the stable orbit condition.
This part of the effective potential will not change when the cosmological constant effects are included or disappear.
In this case, both the critical point and its value depend on the Finslerian parameter, and this difference between the two models are shown in the figure.\\
Figure \ref{fig7} shows the effective potential curve under the condition $L\gg2M$ at scale for scale $ r\approx(\frac{3M}{\Lambda})^{\frac{1}{3}}$. In this case, the obtained new saddle point does not depend on the Finslerian parameter.
Also, Effective potentials in the Finslerian and Riemannian models coincide. If there is a difference, it is very slight and can be ignored.\\
This discrepancy between the two models may be due to Lorentz violations or the minor addition of energy to the gravitational potential of Finsler space-time.\\
Trajectories of space-time play an essential role in studying gravitational waves and black hole shadows.
In future work, We will explore the quintessential influence on shadows of black holes in Finslerian space-time.


\end{document}